\def\la{\;
\raise0.3ex\hbox{$<$\kern-0.75em\raise-1.1ex\hbox{$\sim$}}\; }
\def\ga{\;
\raise0.3ex\hbox{$>$\kern-0.75em\raise-1.1ex\hbox{$\sim$}}\; }
\documentstyle[psfig,11pt]{article}
\begin{document}
\begin{center}
{\Large 
The primordial deuterium abundance~: problems and prospects} 
\end{center}

\medskip\noindent
Sergei A. Levshakov$^{1,2}$, 
Fumio Takahara$^{3}$, and  
Wilhelm H. Kegel$^{4}$ 

\medskip\noindent
{\small
$^{1}$National Astronomical Observatory, Mitaka, Tokyo 181, Japan\\
$^{2}$A. F. Ioffe Physico-Technical Institute, 194021 St.Petersburg, Russia \\
$^{3}$Department of Earth and Space Science, Faculty of Science,
Osaka University, Toyonaka,\\ Osaka 560, Japan\\
$^{4}$Institut f\"ur Theoretische Physik der Universit\"at Frankfurt am Main,
Postfach 11 19 32,\\ 60054 Frankfurt/Main 11, Germany}

\bigskip
\begin{center}
{\large ABSTRACT}
\end{center}

\bigskip\noindent

The current status of extragalactic deuterium abundance 
is discussed using two examples of `low' and `high' D/H measurements.
We show that the discordance of these two types of D abundances
may be a consequence of the spatial correlations in the
stochastic velocity field.
Within the framework of the
generalized procedure (accounting for such effects)
one finds good agreement between 
different observations
and the theoretical predictions for 
standard big bang nucleosynthesis (SBBN).
In particular, we show
that the deuterium absorption seen at $z_a = 2.504$ toward 
Q~1009+2956 and 
the H+D Ly$\alpha$ profile observed at $z_a = 0.701$ toward
Q~1718+4807 are compatible with 
D/H  $\sim 4.1 - 4.6\times10^{-5}$.
This result supports SBBN and, thus, no inhomogeneity is needed.
The problem of precise D/H measurements is discussed.

\begin{center}
{\bf Introduction}
\end{center}

The measurement of deuterium abundance at high redshift from
absorption spectra of QSOs is the most sensitive test of physical
conditions in the early universe just after the era of big bang
nucleosynthesis. The standard BBN model predicts  strong
dependence of the primordial ratio of deuterium to hydrogen nuclei
D/H on the cosmological baryon-to-photon ratio $\eta$
and the effective number of light neutrino species $N_\nu$
(e.g. Sarkar 1996). 
According to the basic idea of homogeneity and isotropy of big bang theory
the {\it primordial} deuterium abundance should not vary
in space. One can only expect that the D/H ratio 
decreases with cosmic time 
due to conversion of D into $^3$He and heavier elements in stars.

On the other hand, current measurements of D abundance at high redshift
reveal a scatter of the D/H values over approximately one order of magnitude.
For instance, recent HST observations of the
$z_a = 0.701$ absorption-line system toward the quasar Q~1718+4807
(Webb {\it et al.}, 1997a,b) show D/H =$1.8-3.1\times10^{-4}$. 
This hydrogen isotopic ratio is
significantly higher than that derived from other quasar spectra at
$z_a = 2.504$ [D/H =$1.8-3.5\times10^{-5}$ by Burles \& Tytler, 1996;
D/H = $2.9-4.6\times10^{-5}$ by Levshakov, Kegel \& Takahara, 
1997 (LKT, hereinafter)],
and at $z_a = 3.572$ [D/H = $1.7-2.9\times10^{-5}$ by Tytler {\it et al.},
1996; D/H $ > 4\times10^{-5}$ by Songaila {\it et al.}, 1997].

The apparent spread of the D/H values makes some authors to assume
fluctuations in the baryon-to-photon ratio at the epoch of BBN,
i.e. to assume that big-bang nucleosynthesis has occurred inhomogeneously
(see e.g. Webb {\it et al.}, and references cited therein).

Current modeling of the evolution of D/H does not allow to estimate
the primordial deuterium abundance with a sufficient accuracy.
The depletion factor is strongly model dependent and changes
from about 2--3 for the chemical evolution models assuming inflow
(e.g. Prantzos 1996,  Tosi {\it et al.} 1997) up to 10 for the models
employing a galactic wind, i.e. outflow (e.g. Scully {\it et al.} 1997).
This means that both `high' ($\sim 2\times10^{-4}$) and `low'
($\sim 3\times10^{-5}$) D/H ratios do not contradict the SBBN prediction
if the ISM deuterium abundance of $1.6\times10^{-5}$
(e.g. Linsky \& Wood, 1996) is adopted. Nevertheless, 
the precise measurements
of absolute values of D/H at high redshift are extremely
important to check whether BBN does occur
inhomogeneously or 
the basic idea of homogeneity and isotropy is still valid.
The fundamental character of this cosmological test requires 
an unambiguous interpretation of spectral observations.

\begin{center}
{\bf How accurately can be determined D/H ?}
\end{center}

It is well known that the physical parameters  derived from 
spectral data depend on the assumptions made with respect to
the line broadening mechanism. For interstellar (intergalactic)
absorption lines a `non-thermal broadening' is usually assumed
to be caused by a large scale motion of the absorbing gas.
The commonly used microturbulent approach considers bulk motions
being completely uncorrelated. This yields a symmetrical (Gaussian)
distribution of the velocity components parallel to the line of
sight, which in turn leads to a symmetrical line profile function being
a convolution of a Voigt profile and a Gaussian. 

Actually, any turbulent flow exhibits an immanent structure in which the
velocities in neighboring volume elements are correlated with each other.
Different aspects
of the line formation processes in correlated turbulent media have been
recently discussed in a series of papers by Levshakov \& Kegel 
(1997, LK hereinafter),
Levshakov, Kegel \& Mazets 
(1997, LKM hereinafter), and by LKT.

Once the statistical properties of a turbulent cloud have been specified,
the way spectral lines ought to be calculated, depends on the problem
considered. Considering emission lines one is dealing with many lines
of sight and the observed intensity should closely correspond to the theoretical
expectation value (see e.g. Albrecht \& Kegel 1987).
If, however, one observes the absorption spectrum in 
the light of a point-like background source, one is dealing with one line of
sight only, and the actually observed intensity may deviate substantially
from the expectation value (for details see LK and LKM), since averaging 
along one line of sight only, corresponds to averaging over an incomplete
sample. The observed line profile is determined by the velocity distribution
along the particular line of sight. For large values of the
ratio of the cloud thickness $L$ to the correlation length $l$ 
the distribution function $p(v)$ for the velocity component along the
line of sight approaches the statistical average, which has been assumed
to be a Gaussian. For values of $L/l$ of only a few, however, $p(v)$ may
deviate substantially from a Gaussian, and is asymmetric in general.
This leads to  
a complex shape of the absorption coefficient for which
the assumption of Voigt profiles could be extremely misleading.
The actual D/H ratio may turn out to be {\it higher} or {\it lower} than the
value obtained from the Voigt-fitting procedure.

The present paper is primarily aimed at the inverse problem
in the analysis of the H+D Ly$\alpha$ absorption observed by Burles \& Tytler (1996) and by
Webb {\it et al.} (1997a,b). The original analysis was performed in
the framework of the microturbulent model. Here we  make an attempt
to re-analyze the observational data on the basis of a more general 
mesoturbulent model. We consider a cloud with a turbulent velocity
field with finite correlation length but of homogeneous (HI-)density and
temperature. The velocity field is characterized by its rms amplitude  $\sigma_t$
and its correlation length $l$.
The model is identical to that of LKT. --  The objective is to 
investigate whether the data in question may be interpreted by 
a unique D/H ratio consistent with the primordial abundances of $^4$He  and
$^7$Li and theoretical SBBN predictions.
 

\begin{center}
{\bf Parameter estimation by the RMC procedure}
\end{center}

To estimate physical parameters and an appropriate velocity field
structure along the line of sight, we used a Reverse Monte Carlo [RMC]
technique, i.e. a stochastic optimization algorithm 
developed to solve optimization problems with a very large number
of free parameters (see LKT). 

The algorithm requires to define 
a simulation box for the 5 physical parameters~:
N(HI), D/H, $T_{kin}$, $\sigma_t/v_{th}$, and $L/l$
(here $v_th$ denotes the thermal width of hydrogen lines).
 -- The continuous random function of the coordinate $v(s)$ is represented by
its sampled values at equal space intervals $\Delta s$, i.e. by
$\{v_1, v_2, \dots , v_k\}$, the
vector of the velocity components parallel to the line of sight
at the spatial points $s_j$. The total number of intervals depends on
the values of $\sigma_t/v_th$ and $L/l$, being typically $\sim 100$
for hydrogen absorption lines.

The convergence of the RMC procedure depends on the size of the simulation box
which is unknown a priori. If it is too large, the computing time increases
considerably, and if it is too small, the required minimum of the objective
function may not be reached at all. Therefore the box size must be chosen
with some care and should be adjusted to the experimental data.
In the next sections we consider two examples of the RMC application.

\begin{center}
{\bf Deuterium abundance at $z = 2.504$ toward Q 1009+2956}
\end{center}

We applied at first the RMC procedure to a 
template H+D Ly$\alpha$ profile which reproduces the
Q1009+2956 spectrum with the DI Ly$\alpha$ line seen at $z_a = 2.504$ (Burles \& Tytler 1996).
Here we present only a part of our results, the full analysis is given by LKT.
\begin{figure*}
\vspace{2.2cm}
\hspace{1.6cm}\psfig{figure=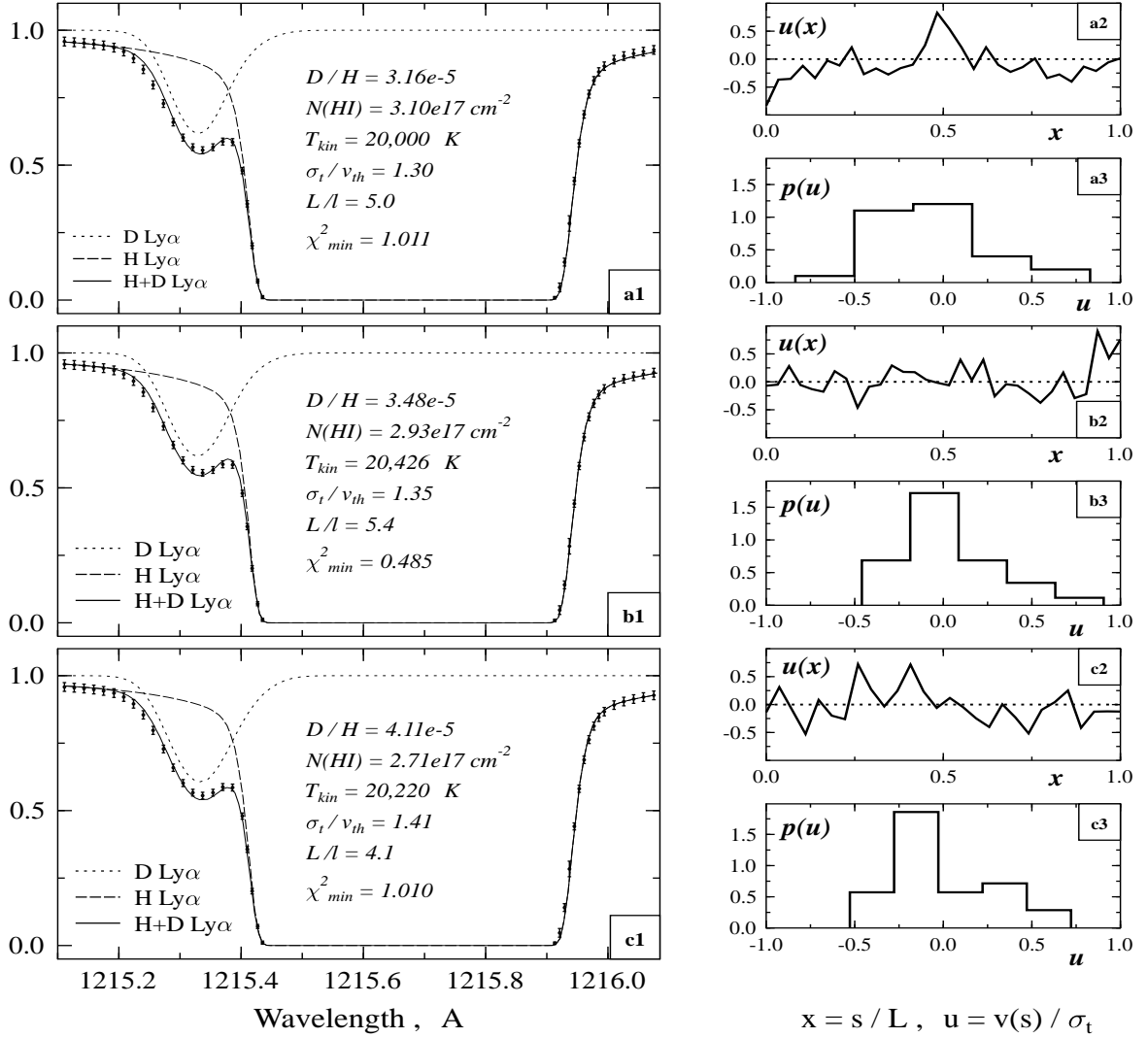,height=12.0cm,width=13.0cm}
\vspace{-0.7cm}
\caption[]{
({\bf a1}, {\bf b1}, {\bf c1}) - 
A template H+D Ly$\alpha$ profile (dots with error bars) 
representing the normalized intensities and
their uncertainties of the $z = 2.504$ system toward Q~1009+2956 
in accord with Burles \& Tytler (1996). The solid curves
show the results of the RMC minimization, whereas the dotted and 
dashed curves are the separate profiles of DI and HI, respectively. 
Also shown are the best fitting parameters and the $\chi^2$ 
values obtained for each case.\\
({\bf a2}, {\bf b2}, {\bf c2}) - 
The corresponding individual realizations of the velocity 
distribution $u(x)$ in units of $\sigma_t$.\\
({\bf a3}, {\bf b3}, {\bf c3}) - 
The histograms are the projected velocity distributions $p(u)$.
}
\end{figure*}

To simulate real data, we added the experimental uncertainties to the
template intensities which were sampled in equidistant bins as shown in Fig.~1
by dots and corresponding error bars. {\it One}-component mesoturbulent model
with a {\it homogeneous} density and temperature was adopted. Adequate profile fits
for three different sets of parameters are shown in panels
({\bf a1}, {\bf b1}, {\bf c1}) by solid curves, whereas the individual HI and DI
profiles are the dashed and dotted curves, respectively. The estimated parameters
(D/H, $N_{\rm HI}$, $T_{kin}$, $\sigma_t / v_{th}$, $L/l$) and $\chi^2_{min}$ values 
per degree of freedom are
also listed in these panels for each RMC solution.

The essential difference between the results of the {\it two}-component
microturbulent model adopted by Burles \& Tytler and ours lies in the estimation
of the hydrodynamical velocities in the $z_a = 2.504$ absorbing region.
It is generally believed that absorption line systems with
N(HI)$ \sim 10^{17}$ cm$^{-2}$ arise in the halos of putative
intervening galaxies (e.g. Tytler {\it et al.} 1996). 
The RMC procedure yields $\sigma_t \simeq 25$ km s$^{-1}$, whereas the
{\it two}-component model leads to $\sigma_t \simeq 2$ km s$^{-1}$ which
is evidently too low as compared with direct observations
of galactic halos at $z > 2$ (van Ojik {\it et al.} 1997) which show that 
$\sigma_t \simeq 40 \pm 15$ km s$^{-1}$,  if $T_{kin} \simeq 10^4$ K.

In this particular absorption system, 
the study of the H+D Ly$\alpha$ profile yields D/H = $(3.75 \pm 0.85)\times10^{-5}$
which is slightly higher than the value $2.51^{+0.96}_{-0.69} \times 10^{-5}$
found by Burles \& Tytler. This difference is, however, very significant because it
leads to limits on D/H consistent with other observations discussed below.

\begin{center}
{\bf Deuterium abundance at $z = 0.701$ toward Q 1718+4807}
\end{center}

The absorbing material at $z = 0.701$ provides an  accurate determination of
the total hydrogen column density due to the extreme sharpness of the Lyman break
measured by the International Ultraviolet Explorer (IUE) satellite (see Webb
{\it et al.}). 

To analyze this system we adopted for
the physical parameters the following boundaries (details are given in Levshakov, Kegel \&
Takahara 1998)~: 

\smallskip\noindent
1.70$\times10^{17}$ cm$^{-2} \leq$ N(HI) $\leq 1.78\times10^{17}$ cm$^{-2}$

\smallskip\noindent
For D/H we use the range from 3.0$\times10^{-5}$ 
to 5.0$\times10^{-5}$, trying to find 
a low D/H solution.

\smallskip\noindent
For $T_{kin}$ we use the interval $10^4 - 2\times10^4$ K, and, thus,
$\sigma_t/v_{th}$ may range within 1.3 -- 4.3.

\smallskip\noindent
For $L/l \gg 1$ the meso- and microturbulent profiles tend to be identical
(see LK). 
Different microturbulent models have been 
thoroughly investigated by Webb {\it et al.},
therefore we will consider moderate 
$L/l$ ratios in the range 1.0 -- 5.0.
\begin{figure*}
\vspace{2.7cm}
\hspace{1.6cm}\psfig{figure=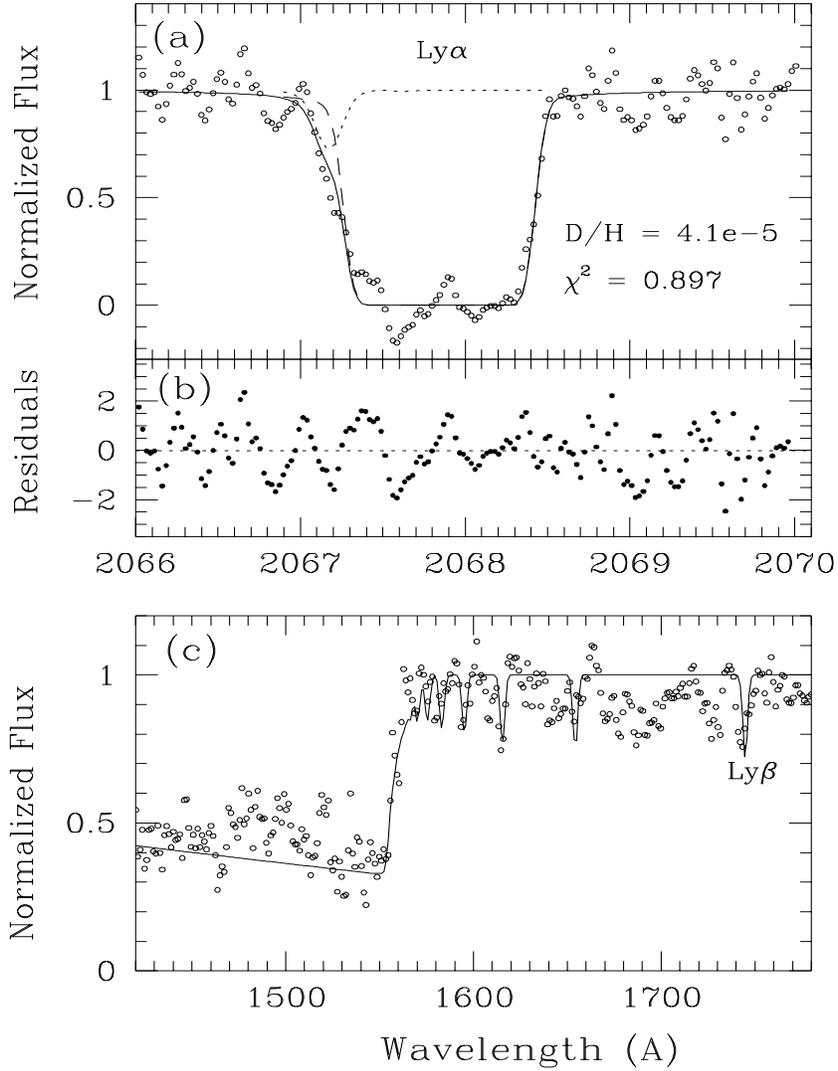,height=12.0cm,width=13.0cm}
\vspace{-0.7cm}
\caption[]{
Observations and RMC fits for Q~1718+4807. 
({\bf a}) -- HST/GHRS data
(open circles) and calculated profiles for HI (dashed curve), 
DI (dotted curve) and H+D (solid curve). The spectral resolution 
corresponds to a FWHM = 0.1 \AA. Shown is an acceptable
solution of the inverse problem with D/H = $4.111\times10^{-5}$,
N(HI) = $1.768\times10^{17}$ cm$^{-2}$, $T_{kin} = 1.51\times10^4$ K,
$\sigma_t = 26$ km s$^{-1}$, and $L/l = 3.5$ (see Table~1).
({\bf b}) -- residuals $\epsilon$ in units of $\sigma_{noise}$ 
(see text).\\
({\bf c}) -- IUE spectrum (open circles) 
and fit (solid curve) with a spectral
resolution corresponding to a FWHM = 2.95 \AA. 
}
\end{figure*}

Following Webb {\it et al.} (1997b), we exclude the Si III line
from our analysis of the H+D Ly$\alpha$ profile. 
As shown by Vidal-Madjar {\it et al.} (1996), `deducting lines
of sight velocity structure for D/H evaluations from ionized species
could be extremely misleading'. 
But we fix $z_a({\rm SiIII}) = 0.701024$
as a more or less arbitrary reference 
radial velocity at which $v_j = 0$.
\begin{figure*}
\vspace{2.7cm}
\hspace{1.6cm}\psfig{figure=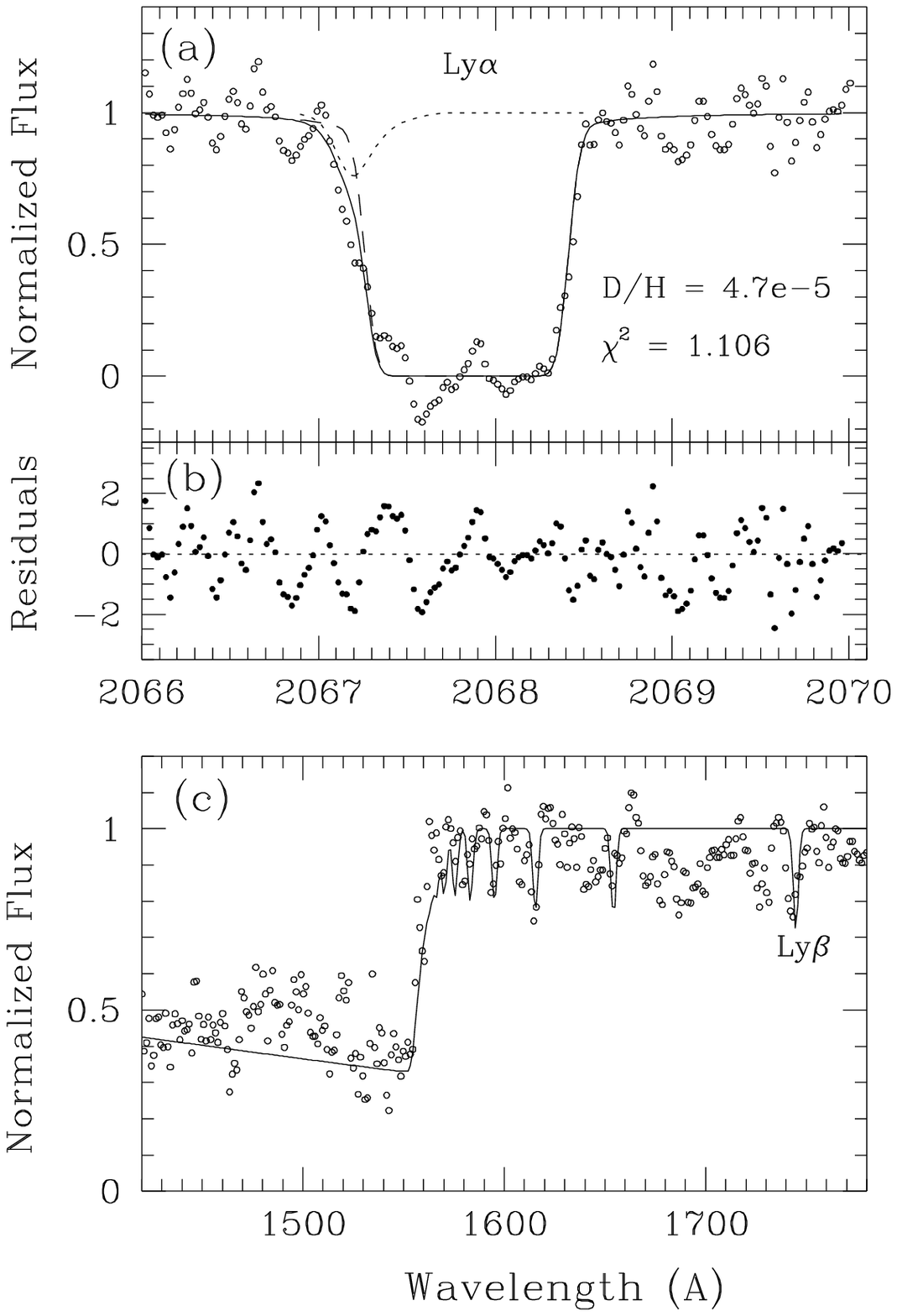,height=12.0cm,width=13.0cm}
\vspace{-0.7cm}
\caption[]{
As Fig. 2 but for D/H = $4.755\times10^{-5}$,
N(HI) = $1.759\times10^{17}$ cm$^{-2}$, $T_{kin} = 1.75\times10^4$ K,
$\sigma_t = 40$ km s$^{-1}$, and $L/l = 2.8$ (see Table~1).
}
\end{figure*}

The estimated parameters for a few  adequate RMC profile fits 
are listed in Table~1.
The derived deuterium abundance 
$\langle {\rm D/H} \rangle \simeq 4.4\times10^{-5}$ is about 4--7 times
smaller than the limiting values of $1.8 - 3.1\times10^{-4}$ found
by Webb {\it et al.} (1997b) in the microturbulent model excluding the Si III line. 
To illustrate our results, we show in Figs. 2 and 3
H+D Ly$\alpha$ profiles for the two calculations
with the lowest and the highest D abundances found in the mesoturbulent model
(D/H = $4.111\times10^{-5}$ and $4.755\times10^{-5}$, respectively).
They are shown
by the solid curve, whereas the open circles give  
the experimental intensities. 
The residuals 
shown in Figs. 2{\bf b} and 3{\bf b}
by filled circles are normally distributed with
zero mean and unit variance.
\begin{figure*}
\vspace{2.7cm}
\hspace{5.4cm}\psfig{figure=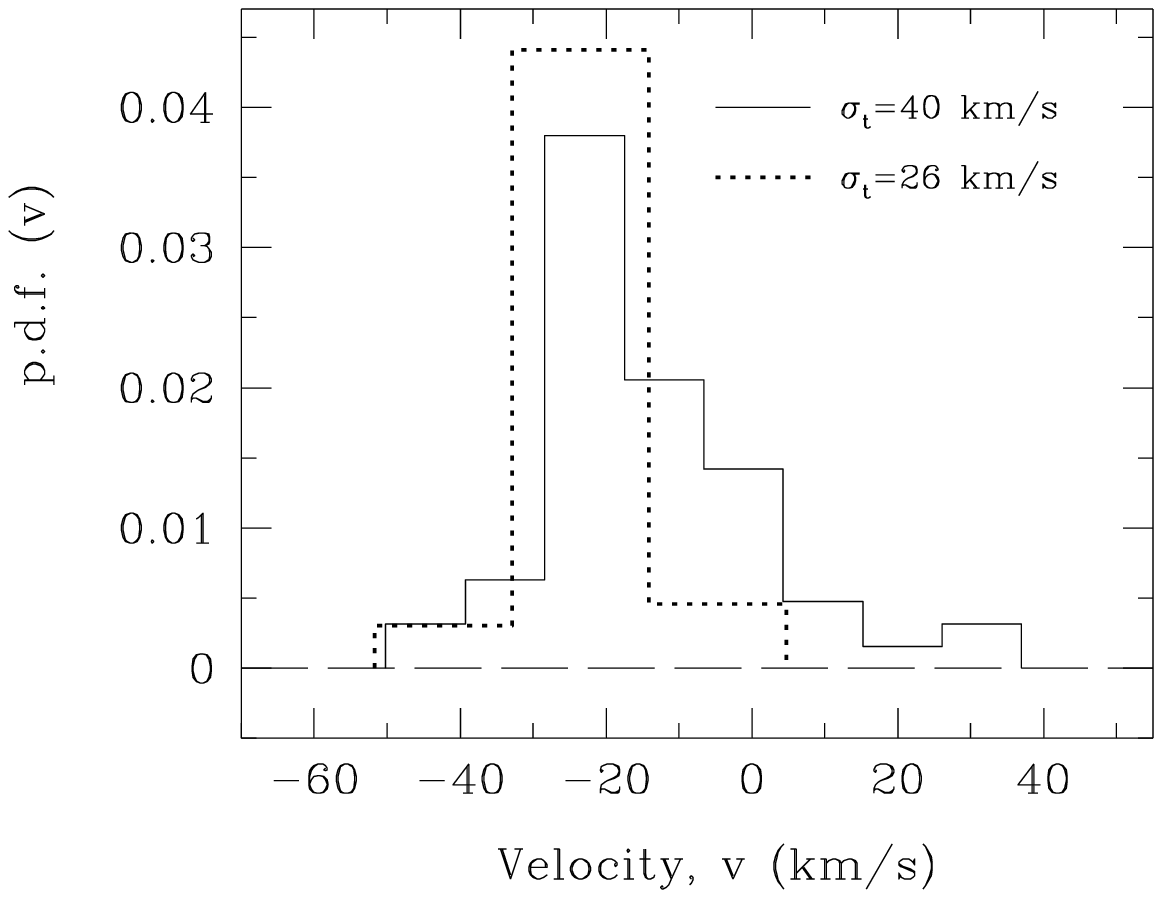,height=6.5cm,width=7.0cm}
\vspace{-2cm}
\caption[]{
Probability density functions $p(v)$ for the
velocity components parallel to the line of sight for two acceptable 
RMC solutions shown in Fig.2{\bf a} (dotted line histogram) 
and in Fig.3{\bf a} (solid line histogram). Both are blue-shifted 
by about $-20$ km s$^{-1}$
with respect to $z_a({\rm SiIII})$ and have asymmetric shapes. 
}
\end{figure*}

To control the $v(s)$-configurations 
estimated  by the RMC procedure,
we calculated profiles for the higher
order Lyman lines and the shape of the Lyman-limit 
discontinuity and then superposed them to the corresponding part of the 
Q~1718+4807 spectrum. 
One expects to find a displacement between the observed and 
the calculated steep Lyman edge 
if the velocity field structure is far from the truth.
The results are shown in Figs. 2{\bf c} and 3{\bf c}
where again open circles correspond to the observed intensities and the computed
spectra are shown by the solid curves. We do not find any pronounced discordance
of calculated and real spectra. 
On the other hand, the spectral resolution of 2.95 \AA\ (FWHM) is not 
sufficient to follow a fine velocity field structure within the
$z_a = 0.701$ absorber. 

The derived $v(s)$-configurations are not unique. Table~1 demonstrates the
spread of the apparent rms turbulent velocities which range from 
18 up to 40 km s$^{-1}$. 
It is worthwhile to emphasize once more that the projected velocity
distribution function (p.d.f.) may differ considerably from  a
Gaussian. Fig.~4 shows an example of such
distortions caused by a poor statistical sample 
(i.e. incomplete averaging)
of the velocity field 
distributions for the two cases of the lowest and the highest D/H ratios 
from Table~1
($\sigma_t = 26$ and 40 km s$^{-1}$, respectively).
Both $p(v)$ distributions are asymmetric.
This is the main reason why the absorption in 
the blue wing of the HI Ly$\alpha$ line may be enhanced without any
additional HI interloper(s). 

\begin{table}
\centering
\caption{Cloud parameters derived from the Ly$\alpha$
profile by the RMC method}
\begin{tabular}{ccccccc}
\hline\noalign{\smallskip}
N(HI) &  D/H & $T_{kin}$  & 
$\sigma_t/v^{\rm H}_{th}$ & $\sigma_t$ &
$L/l$ & $\frac{1}{\nu}\chi^2_{min}$ \\ 
\noalign{\smallskip}
[$10^{17}$ cm$^{-2}$] &  & [$10^4$ K] &  & [km s$^{-1}$] &  &  \\
\noalign{\smallskip}
\hline
\noalign{\smallskip}
1.732  & 4.565 & 1.41 & 1.40 & 22 & 2.7 & 1.064 \\
1.739  & 4.562 & 1.60 & 1.10 & 18 & 3.9 & 1.002 \\
1.759  & 4.755 & 1.75 & 2.33 & 40 & 2.8 & 1.106 \\
1.761  & 4.555 & 1.46 & 1.86 & 29 & 4.3 & 1.086 \\
1.768  & 4.111 & 1.51 & 1.61 & 26 & 3.5 & 0.897 \\
1.771  & 4.442 & 1.76 & 1.34 & 23 & 3.4 & 1.162 \\
1.776  & 4.249 & 1.62 & 1.71 & 28 & 4.0 & 1.114 \\ 
\noalign{\smallskip}
\hline
\end{tabular}
\end{table}

\begin{center}
{\bf How to test the validity of the RMC results ?}
\end{center}

We show that the H+D measurements from two absorption-line systems do not require
either a low or a high value of D/H. 
Both systems at $z = 0.701$ toward Q~1718+4807 and at $z = 2.504$ toward Q~1009+2956 
may be modeled within the framework of the generalized radiative transfer theory
with D/H in the range $4.1 - 4.6\times10^{-5}$.
The fitting procedure leads in this case, however, to asymmetric
velocity field distributions. This fact may be used to test the RMC results
by additional observations of higher order Lyman lines.

Indeed, if $p(v)$ is asymmetric, this will show up in the profile shapes
of the higher order Lyman lines. For example, for the physical parameters listed in
Table~1, the effect becomes visible starting from Ly-4 
[the Ly-$\alpha$, -$\beta$, -$\gamma$ lines are insensitive to the
asymmetry of $p(v)$ due to their high optical depth].
Fig.~5 shows simulated spectra (convolved with
a Gaussian instrumental profile of FWHM = 0.1 \AA) for Ly-4 and Ly-12 using
the same $p(v)$ distributions depicted in Fig.~4 -- dotted curves for
$\sigma_t = 26$ km s$^{-1}$, and solid curves for
$\sigma_t = 40$ km s$^{-1}$. As seen, line shapes may vary depending on
the velocity field structure that makes them asymmetric in general.
As for microturbulent profiles shown in Fig.~5 by the short dash curves, 
they remain symmetrical. The higher order Lyman lines 
may be used unambiguously to constrain the velocity field structure of the
absorption system. Therefore high resolution observations 
of the whole Lyman series become crucial for the D/H measurements.

\begin{figure*}
\vspace{0.0cm}
\hspace{5.4cm}\psfig{figure=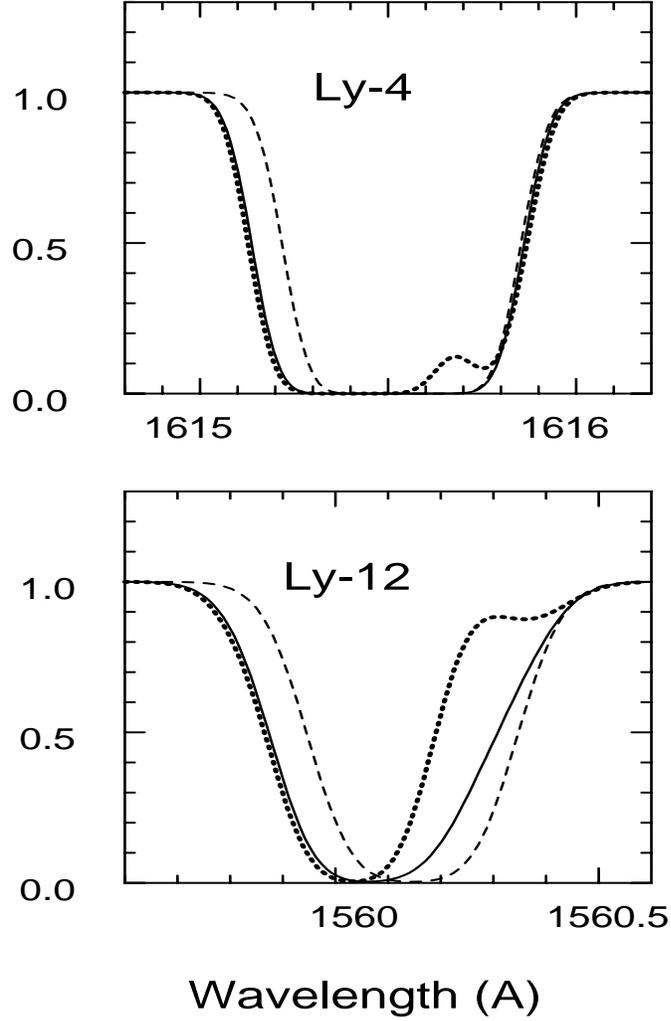,height=12.5cm,width=14.0cm}
\vspace{-1cm}
\caption[]{
The HI Ly-4 and Ly-12 mesoturbulent spectra for 
$\sigma_t = 40$ km s$^{-1}$ (solid curves)
and $\sigma_t = 26$ km s$^{-1}$ (dotted curves).
The corresponding physical parameters are listed in Table~1. 
Short dash curves show microturbulent profiles 
calculated for the mean 
N(HI) = $1.738\times10^{17}$ cm$^{-2}$ and the mean Doppler 
parameter $b$(HI) = 25.5 km s$^{-1}$
of the data by Webb {\it et al.}
The spectra are convolved with a Gaussian instrumental 
profile of FWHM = 0.1 \AA. 
}
\end{figure*}

\begin{center}
{\bf Comparison of the primordial abundances of the light elements}
\end{center}

The revealed mean value of 
$\langle {\rm  D/H} \rangle \simeq 4.4\times10^{-5}$ 
corresponds to the mass fraction of $^{4}$He $Y_p \simeq
0.245$, the $^{7}$Li primordial abundance of $\simeq 2.1\times10^{-10}$, 
and $\eta \simeq 4.4\times10^{-10}$. 

The estimated $Y_p$ and $^{7}$Li/H values
lie within the measured regions $0.240 \leq Y_p \leq 0.246$
(Izotov, Thuan, Lipovetsky 1997) and $1.5\times10^{-10} \leq$
$^{7}$Li/H $\leq 2.0\times10^{-10}$ (Bonifacio \& Molaro 1997).
The measurements of $Y_p$ may contain, however, some systematic uncertainties.
For instance,  Fields {\it et al.} (1996) and Olive {\it et al.} (1997) 
give the value $Y_p = 0.234 \pm 0.003({\rm stat.}) \pm 0.005({\rm syst.})$
and note that deducing abundances from linestrengths under the usual assumptions
about extragalactic HII regions `could introduce a significant systematic
error', the magnitude and the distribution of which are `not well understood'.

The above baryon-to-photon ratio yields the current baryon density
$\Omega_b \simeq 0.03$, assuming a Hubble constant ${\rm H}_0 = 70$
km s$^{-1}$ Mpc$^{-1}$ and the temperature of the cosmic
background radiation $T_{CBR} = 2.726$ K. 
This value of $\Omega_b$ being larger than the observed density
of baryons in stars and hot gas $\Omega_{b,{\rm obs}} \simeq 0.01$
(Fukugita {\it et al.} 1996) implies that most baryons are in
the hot intergalactic gas.

\begin{center}
{\bf Conclusion}
\end{center}

It is not unlikely that the same spectra may yield different
column density estimations. Model dependence of the D/H measurements
has been demonstrated within the framework of microturbulent models
by Wampler (1996) and on the basis of mesoturbulent models 
in a series of our papers (see LKT and references therein).
Macroscopic motions in QSO absorption systems have been repeatedly
confirmed by direct observations (e.g. Songaila {\it et al.} 1995).
Our analysis shows that such large scale motions
may affect the velocity field configuration along the line of sight and the
relative positions of the DI and HI lines (Levshakov \& Takahara 1996) 
making their
unambiguous interpretation difficult. We stress again that the
reliability of the interpretation of the D/H measurements is determined by two factors~:  
($a$) improvements in the detection equipment and ($b$)  advances in the theory of line
formation in turbulent media.

\begin{center}
{\bf Acknowledgments}
\end{center}

\medskip\noindent
The authors are grateful to John Webb for making available the
calibrated IUE and 
HST/GHRS spectra of Q~1718+4807 and acknowledge helpful correspondence
and comments by him and Alfred Vidal-Madjar. 
This work was supported in part by the RFBR grant No. 96-02-16905a.

\begin{center}
{\bf References }
\end{center}

\medskip\noindent
Albrech M. A., Kegel W. H., 1987, A\&A, 176, 317\\
Bonifacio P., Molaro P., 1997, MNRAS, 285, 847\\
Burles S., Tytler D., 1996, astro-ph/9603069\\
Fields B. D., Kainulainen K., Olive K. A., Thomas D., 1996,
New Astronomy, 1, 77\\
Fukugita M., Hogan C. J., Peebles P. J. E., 1996, Nature, 381, 489\\
Izotov Yu., Thuan T. X., Lipovetsky V. A., 1997, ApJS, 108, 1\\
Levshakov S. A., Takahara F., 1996, Astron. Letters, 22, 438\\
Levshakov S. A., Kegel W. H., 1997, MNRAS, 288, 787 [LK]\\
Levshakov S. A., Kegel W. H., Mazets I. E., 1997, MNRAS, 288, 802 [LKM]\\
Levshakov S. A., Kegel W. H., Takahara F., 1997, MNRAS ({\it submit.}),
astro-ph/9710122 [LKT]\\
Levshakov S. A., Kegel W. H., Takahara F., 1998, A\&A ({\it submit.})\\
Linsky J., Wood B. E., 1996, ApJ, 463, 254\\
Olive K. A., Steigman G., Skillman E. D., 1997, ApJ, 483, 788\\
Prantzos N., 1996, A\&A, 310, 106\\
Sarkar S., 1996, Rep. Prog. Phys., 59, 1493\
Scully S., Cass\'e M., Olive K. A., Vangioni-Flam E., 1997, ApJ, 476, 521\\
Songaila A., Hu E. M., Cowie L. L., 1995, Nature, 375, 124\\
Songaila A., Wampler E. J., Cowie L. L., 1997, Nature, 385, 137\\
Tosi M., Steigman G., Matteucci F., Chiappini C., 1997, 
ApJ ({\it submit.}), astro-ph/9706114\\
Tytler D., Fan X.-M., Burles S., 1996, Nature, 381, 207\\
Vidal-Madjar A., Ferlet R., Lemoine M., 1996, astro-ph/9612020\\
Wampler E. J., 1996, Nature, 383, 308\\
Webb J. K., Carswell R. F., Lanzetta K. M., Ferlet R., Lemoine M.,\\
\hspace*{1cm}Vidal-Madjar A., Bowen D. V., 1997a, Nature, 388, 250\\
Webb J. K., Carswell R. F., Lanzetta K. M., Ferlet R., Lemoine M.,\\
\hspace*{1cm}Vidal-Madjar A., 1997b, astro-ph/9710089

\end{document}